\documentclass[preprint,showpacs,preprintnumbers,amsmath,amssymb,superscriptaddress]{revtex4}

\usepackage{chemformula} 
\usepackage[T1]{fontenc} 
\usepackage{amssymb}
\usepackage{amsmath}

\usepackage{hyperref}

\newcommand*\VEC[1]{\boldsymbol{#1}}

\begin{document}

\author{Ping-Yuan Lo}\thanks{P. Y. Lo and G. H. Peng contribute equally to this work and their names are listed by alphabetical order.}
\affiliation{Department of Electrophysics, National Chiao Tung University, Hsinchu 30010, Taiwan}
\affiliation{Department of Electrophysics, National Yang Ming Chiao Tung University, Hsinchu 30010, Taiwan}

\author{Guan-Hao Peng}\thanks{P. Y. Lo and G. H. Peng contribute equally to this work and their names are listed by alphabetical order.}
\affiliation{Department of Electrophysics, National Chiao Tung University, Hsinchu 30010, Taiwan}
\affiliation{Department of Electrophysics, National Yang Ming Chiao Tung University, Hsinchu 30010, Taiwan}

\author{Wei-Hua Li}
\affiliation{Department of Electrophysics, National Chiao Tung University, Hsinchu 30010, Taiwan}
\affiliation{Department of Electrophysics, National Yang Ming Chiao Tung University, Hsinchu 30010, Taiwan}

\author{Yi Yang}
\affiliation{Department of Electrophysics, National Chiao Tung University, Hsinchu 30010, Taiwan}
\affiliation{Department of Electrophysics, National Yang Ming Chiao Tung University, Hsinchu 30010, Taiwan}

\author{Shun-Jen Cheng}
\affiliation{Department of Electrophysics, National Chiao Tung University, Hsinchu 30010, Taiwan}
\affiliation{Department of Electrophysics, National Yang Ming Chiao Tung University, Hsinchu 30010, Taiwan}
\email{sjcheng@mail.nctu.edu.tw}

\title{Inherently high valley polarizations of momentum-forbidden dark excitons in transition-metal dichalcogenide monolayers}







\begin{abstract}
High degree of valley polarization of optically active excitons in transition-metal dichalcogenide monolayers (TMD-MLs) is vital in valley-based photonic applications but known to be likely spoiled by the intrinsic electron-hole exchange interactions.
In this study, we present a theoretical investigation of the valley and optical properties of finite-momentum dark excitons in WSe$_2$-MLs by solving the density-functional-theory(DFT)-based Bethe-Salpeter equation (BSE) under the guidance of symmetry analysis.
We reveal that, unlike the bright exciton inevitably subjected to electron-hole exchange interaction, inter-valley finite-momentum dark excitons in WSe$_2$-MLs are well immune from  the exchange-induced valley depolarization and inherently highly valley-polarized under the enforcement of the crystal symmetries.
More importantly, the superior valley polarizations of the inter-valley dark excitons in WSe$_2$-MLs are shown almost fully transferable to the optical polarization in the phonon-assisted photo-luminescences because of the native suppression of exchange-induced depolarization in the second-order optical processes. The analysis of phonon-assisted photo-luminescences accounts for the recently observed  brightness, high degree of optical polarization and long lifetime of the inter-valley dark exciton states in tungsten-based TMD-MLs.

\end{abstract}

\maketitle


\noindent
\emph{Introduction---} Transition-metal dichalcogenide monolayers (TMD-MLs) have drawn a broad interest in recent years because of the intriguing spin-valley-coupled characteristics in the electronic and excitonic structures. \cite{TFHeinz2010, WYao2012a, WYao2014a}
As a massive Dirac material, the band structure of a TMD-ML is characterized by the inter-band energy gaps opened in the distinct valleys located at the $K$ and $K'$ corners of the first Brillouin zone (BZ) that follow the opposite optical helicity and allow for the valley-selective optical excitation and manipulation.\cite{WYao2014b, WYao2012b, JShan2018a}
With the extraordnary valley-excitonic properties, TMD-MLs serve as promising nano-materials for the application of valley-based photonics as long as the degree of valley polarization of exciton can remain high and well transferrable to the optical polarization of the emitted photons.\cite{GWang2015,Smolenski2016}

However, it is widely known that the valley-polarization of a bright exciton in a TMD-ML is very likely depolarized by the intrinsic electron-hole (\emph{e-h}) exchange interaction that couples the inter-band excitations in the distinct valleys. \cite{WYao2014b, MWWu2014, AHMacDonald2016, GWang2014, Glazov2015, selig2020suppression}
Despite the weak meV-scale coupling strength, the momentum-conserving {\it e-h} exchange interaction (EHEI) can efficiently intermix two exchange-free excitations in the $K$ and $K'$ valleys that hold the {\it same} momentum and similar energy ({\it quasi}-degenerate). \cite{WYao2014b, SGLouie2015a}
Note that, at the most general level, the fundamental time-reversal symmetry (TRS) ensures the degeneracy only for the excitations in the opposite valleys that carry the {\it opposite } momentum. Hence, even without any spatial symmetries, the distinct degenerate valley-exciton states that hold the same {\it nearly vanishing} momentum, i.e. the bright exciton (BX) states, meet the both criteria and natively suffer from the exchange-induced valley depolarization. By contrast, the valley exciton states with the same finite momenta, i.e. momentum-forbidden dark exciton (MFDX) states, are not enforced by the TRS to be degenerate and, unlike the BXs, the valley polarizations of the MFDX states in TMD-MLs are purely dictated by the crystal symmetries and should be momentum-dependent.

In spite of violating the momentum selection rules, those MFDXs in TMD-MLs have drawn massive attention recently because of their essential involvement in various optical and dynamics phenomena. \cite{SFShi2019, CHLui2020, WYao2020a, EMalic2020a, CHLui2019a, bao2020probing, HZhu2019a, Koitzsch2019, KSuenaga2020, EMalic2021a, JZhao2021, EMalic2021b, Simbulan2021}
Very recently,  direct generation and probe of finite-momentum excitons in WSe$_2$-MLs have been realized by using the integrated technology of optical pump-probe and angle-resolved photo-emission spectroscopies \cite{KMDani2020, CMiddleton2021}.
Moreover, recent cryogenic photo-luminescence (PL) measurements on high-quality tungsten-based TMD-ML samples have revealed the pronounced optical signatures of the inter-valley MFDXs that are significantly bright,\cite{CHLui2020} highly polarized\cite{SFShi2019,WYao2020a} and long-lived  \cite{chen2020}. Those long-lived and optically accessible MFDXs are attractive for the quantum applications and realizing excitonic Bose-Einstein condensation (BEC) \cite{KFMak2019a, RRapaport2019, AWHolleitner2020, MCombescot2017}.
With the recently achieved experimental advances in the investigation of MFDXs, it is timely demanded to establish a comprehensive theoretical understanding of those finite-momentum dark excitons in TMD-MLs over the extended momentum space, which yet remains rarely explored so far \cite{Deilmann2019}.

In this Letter, we present a theoretical investigation of the finite-momentum exciton states of WSe$_2$-MLs by numerically solving the DFT-based BSE under the guidance of symmetry analysis.\cite{SJCheng2019} The studies reveal the symmetry-dictated landscape of the momentum-dependent valley and optical properties of the finite-momentum exciton states over the full Brillouin zone and carry out the analysis of phonon-assisted PL, which explains the recently observed spectral brightness \cite{SFShi2019, CHLui2020, WYao2020a}, high degree of optical polarization \cite{SFShi2019,WYao2020a} and long-lived dynamics \cite{chen2020} of the inter-valley MFDXs in tungsten-based TMD-MLs.

\begin{figure}[!ht]
\includegraphics[width=0.9\columnwidth]{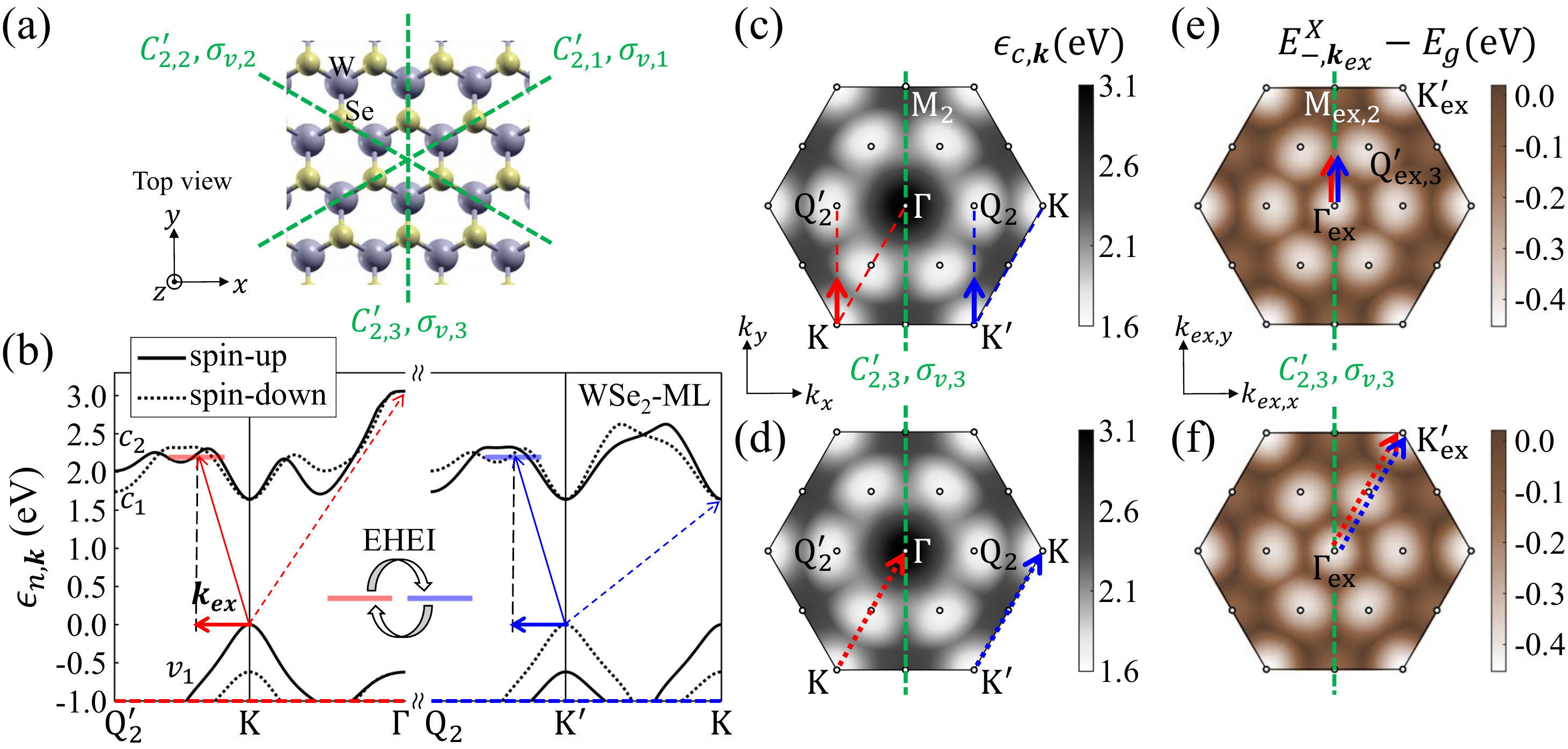}
\caption{(a) Top view of the lattice structure of a TMD-ML with $D_{3h}$ symmetry. The green dashed lines indicate the three axes with respect to the $C_{2}^{'}$ rotational and $\sigma_{v}$ mirror symmetries. (b) The DFT-calculated quasi-particle conduction and valence bands of a WSe$_2$-ML along the specific $\VEC{k}$-paths as indicated by the red and blue dashed lines in the BZ of (c). (c) and (d): Energy contour plot of the DFT-calculated lowest conduction band of a WSe$_2$-ML over the first BZ in the electron-momentum ($\VEC{k}$) space. (e) and (f): The calculated energy of the lowest exciton band of a WSe$_2$-ML over the first BZ in the exciton-momentum ($\VEC{k}_{ex}$) space. 
} 
\label{Fig1}
\end{figure}

\noindent
\emph{Theoretical analysis and numerical methodology---} First, we consider the exciton state with the center-of-mass wave vector $\VEC{k} _{ex}$,
$\left| S , \VEC{k} _{ex} \right\rangle = \frac{1}{\sqrt{\mathcal{A}}} \sum _{v c \VEC{k}} A _{S ,\VEC{k} _{ex}} \!\! \left( v c \VEC{k} \right) \hat{c} _{c, \VEC{k} + \VEC{k} _{ex}} ^{\dagger} \hat{h} _{v, -\VEC{k}} ^{\dagger} | GS \rangle $
, written as a linear combination of the configurations of the electron-hole ({\it e-h} ) pairs, $\hat{c} _{c, \VEC{k} + \VEC{k} _{ex}} ^{\dagger} \hat{h} _{v, -\VEC{k}} ^{\dagger} | GS \rangle $,
where the particle operator $\hat{c}_{c,\VEC{k}}^{\dagger}$ ($\hat{h}_{v,-\VEC{k}}^{\dagger}$) is defined to create the electron (hole) of the wave vector $\VEC{k}$ ($-\VEC{k}$) in the conduction band $c$ (valence band $v$) from the ground state of the system with the fully filled valence bands $| GS \rangle$,
$S$ is the index of exciton band, $A_{S, \VEC{k} _{ex}} \!\! \left( v c \VEC{k} \right)$ is the amplitude of the {\it e-h} configuration $\hat{c} _{c, \VEC{k} + \VEC{k} _{ex}} ^{\dagger} \hat{h} _{v, -\VEC{k}} ^{\dagger} | GS \rangle$ in the exciton state, and $\mathcal{A}$ is the area of the two-dimensional (2D) material.
The exciton wave function in the reciprocal $\VEC{k}$-space, $A _{S, \VEC{k} _{ex}} \!\! \left( v c \VEC{k} \right)$, follows the BSE that reads \cite{LJSham1966, LJSham1980, SGLouie1998, AHMacDonald2015a, Deilmann2019, SJCheng2019, PHawrylak2020, vasconcelos2018dark}
\begin{align}\label{eqn:BSE}
  &\left[ \epsilon _{c, \VEC{k} + \VEC{k} _{ex}} - \epsilon _{v, \VEC{k}} - E _{S, \VEC{k} _{ex}} ^{X} \right] A _{S, \VEC{k} _{ex}} \!\! \left( v c \VEC{k} \right) \notag \\
  &+ \sum _{v ^{\prime} c ^{\prime} \VEC{k} ^{\prime}} U _{\VEC{k} _{ex}} \!\! \left( v c \VEC{k} , v ^{\prime} c ^{\prime} \VEC{k} ^{\prime} \right) A _{S, \VEC{k} _{ex}} \!\! \left( v ^{\prime} c ^{\prime} \VEC{k} ^{\prime} \right) = 0 ,
\end{align}
where $E _{S, \VEC{k}_{ex}} ^{X}$ is the eigen energy of the exciton state, the first two terms on the left hand side are the kinetic energies of free electron and hole, $\epsilon_{c,\VEC{k}+\VEC{k}_{ex}}$, respectively, and $(-\epsilon_{v,\VEC{k}})$, and the last term is associated with the kernel of {\it e-h} Coulomb interaction that consists of the screened {\it e-h} direct interaction and the {\it e-h} exchange one, $U_{\VEC{k} _{ex}}=- V_{\VEC{k} _{ex}} ^{d} + V_{\VEC{k} _{ex}} ^{x}$. 
Figure~\ref{Fig1}b presents the DFT-calculated lowest conduction and topmost valence band of a WSe$_2$-ML along the selected $\VEC{k}$ paths by using the first principles VASP package \cite{GKresse1996} within the Heyd-Scuseria-Ernzerhof (HSE) exchange-correlation functional model. \cite{HSE06} The extended DFT-calculated band structures of a WSe$_2$-ML is presented in Ref.\cite{Supplement}

The explicit definitions of the matrix elements of $V_{\VEC{k} _{ex}} ^{d}$ and $V_{\VEC{k} _{ex}} ^{x}$ in terms of the Bloch wave functions are given in Ref.\cite{Supplement}
Throughout this work, the screening in the direct {\it e-h} Coulomb interaction is modeled by using the Keldysh formalism. \cite{LVKeldysh1979, Supplement, PCudazzo2011, AHMacDonald2015a, TCBerkelbach2013, AVStier2018, ERidolfi2018, MLTrolle2017}
Following the comuptational methodology developed in Ref.~\cite{SJCheng2019}, we establish the BSE in the Wannier tight binding scheme \cite{kosmider2013large,scharf2016excitonic,lado2016landau} based on the DFT-calculated electronic structures \cite{Wannier90a, Wannier90b, Supplement} and solve the exciton states and band structures by means of $\VEC{k}$-grid discretization and direct matrix diagonalization.

It is known that the direct {\it e-h} Coulomb interaction makes the predominant contribution to the large binding energy of exciton in a TMD-ML, but no direct effect on the inter-valley couplings for exciton. \cite{SGLouie2013a, TFHeinz2014a}
By contrast, the meV-scaled EHEIs make the inter-valley excitonic couplings that are especially efficient  between the distinct valley excitons with the same momentum and similar energy (quasi-degeneracy).
As previously stated, the degeneracies of exciton states over the momentum space are dictated by the enforcement the crystal symmetries, so are the degree of valley polarizations of the exciton states. 

Below, we conudct the symmetry analysis to predict the degeneracies of the exchange-free exciton states of TMD-MLs under the $D_{3h}$-group symmetry.\cite{CRobert2017}
Consider two distinct free {\it e-h} pair states with the same $\VEC{k} _{ex}$ excited from the different valence states at $\VEC{k}$ and $\VEC{k} ^{\prime}$. They are degenerate as
$\epsilon_{c, {\VEC{k} + \VEC{k} _{ex}}} - \epsilon _{v, \VEC{k}} = \epsilon _{c, \VEC{k} ^{\prime} + \VEC{k} _{ex}} - \epsilon _{v, \VEC{k} ^{\prime}} $, which can generally hold only if
$\epsilon_{v, \VEC{k}} = \epsilon _{v, \VEC{k} ^{\prime}}$ and $\epsilon _{c, \VEC{k} + \VEC{k} _{ex}} = \epsilon _{c, \VEC{k} ^{\prime} + \VEC{k} _{ex}}$.
From the theory of group representations, the above two equations hold when the space group symmetry of the TMD-ML satisfies the both equations, $\VEC{k} ^{\prime} = \hat{U} \VEC{k}$ and $\VEC{k} ^{\prime} + \VEC{k} _{ex} = \hat{U} \left( \VEC{k} + \VEC{k} _{ex} \right)$, for any symmetry operator $\hat{U} \in D _{3h}=\{E,C_{3},C_{3}^{-1},\sigma_{h},S_{3},S_{3}^{-1},C_{2,1}^{\prime},C_{2,2}^{\prime},C_{2,3}^{\prime},\sigma_{v,1},\sigma_{v,2},\sigma_{v,3}\}$.
Accordingly, we find the criterion for the formation of valley-degeneracy of two distinct {\it e-h} pairs carrying the same $\VEC{k}_{ex}$, i.e.
\begin{equation}
\VEC{k}_{ex} = \hat{U} \VEC{k}_{ex} \, .
\label{exc_deg}
\end{equation}


\begin{figure}[!ht]
\includegraphics[width=0.9\columnwidth]{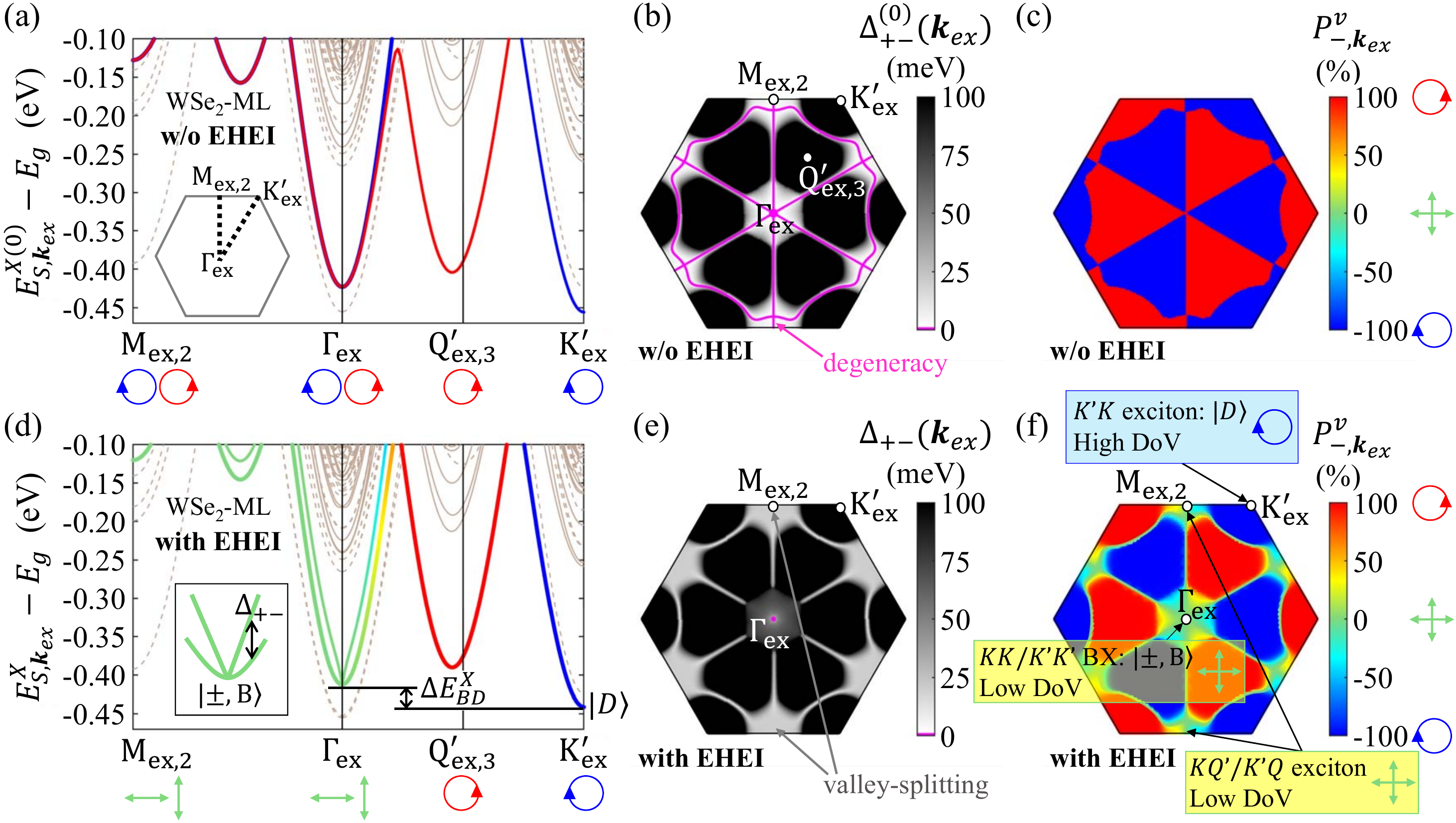}
\caption{ (a) Exciton band structure of a WSe$_2$-ML along the paths of $\overline{\Gamma _{ex} M _{ex,2}}$ and $\overline{\Gamma_{ex} K ^{\prime}_{ex}}$ calculated with the neglect of the {\it e-h} exchange interaction (EHEI). Solid (dashed) lines: spin-like (-unlike) exciton bands. The band colors follow the color-bar scales of (c) and (f) to represent the valley degree of polarizarion of the lowest exciton band.  (b) The $\VEC{k}_{ex}$-dependent energy differences, $\Delta _{+ -} ^{(0)} (\VEC{k} _{ex}) \equiv E_{+,\VEC{k}_{ex}}^{X (0)} - E_{-,\VEC{k}_{ex}}^{X (0)}$, between the lowest spin-like exciton doublet, $|\pm, \VEC{k}_{ex} \rangle $, based on the result of (a). The magenta-coloured line indicates $\Delta_{+-} ^{(0)} (\VEC{k}_{ex})=0$ (the degeneracy of the exciton states)  (c) The $\VEC{k}_{ex}$-dependent valley polarization, $P_{-,\VEC{k}_{ex}}^{v}$, of the lowest spin-like exciton states, $|-,\VEC{k}_{ex} \rangle \equiv |D\rangle $, based on (a). 
(d)-(f) are the same as (a)-(c) but with the inclusion of the {\it e-h} exchange interaction in the calculation. $\Delta E_{BD}^X=29$meV for WSe$_2$-ML.
}	
\label{Fig2}
\end{figure}


\noindent
\emph{Results---} Applying all the symmetry operators $\hat{U} \in D_{3h}$ in Eq.~(\ref{exc_deg}) for all $\VEC{k}_{ex} \in $ BZ,\cite{Supplement} one can show that distinct \emph{e-h} pair states with the common $\VEC{k}_{ex}$ could be valley-degenerate only if $\VEC{k}_{ex}$ lies along the lines connecting the $\Gamma_{ex}$ and $M _{ex,i}$ points, i.e. the axes associated with the $3\sigma_v$ and $3C_2'$ symmetries. 
This predicts the exchange-induced valley depolarization that impacts only the exciton states with the specific $\VEC{k} _{ex} \in \overline{\Gamma_{ex}M_{ex,i}}$, including the commonly known bright exciton around the $\Gamma_{ex}$ point. \cite{MWWu2014, AHMacDonald2016}

As an illustrative instance, Fig.~\ref{Fig1}c exemplifies the two distinctive {\it e-h} pair excitations with the same $\VEC{k}_{ex}$ along the $k_y$-direction, i.e. $\overline{\Gamma_{ex} M_{ex,2}}$, which are excited from the valence $K$ towards the conduction $Q'_{2}$ valley (red arrow line) and from the valence $K'$ towards the conduction $Q_{2}$ valley (blue arrow line), respectively. In Fig.~\ref{Fig1} b, one can identify the transition energies of the two free valley-excitations to be the same.
For comparative illustration, we consider another set of two inter-band transitions excited from the distinctive valence valleys with the common $\VEC{k}_{ex}$ along $\overline{\Gamma_{ex} K_{ex}'}$, as depicted in Fig.~\ref{Fig1}d. With the misaligned $\VEC{k}_{ex}$ from $\overline{\Gamma_{ex} M_{ex,i}}$  the transition energies  are  apparently different as predicted by analysis and seen in Figure~\ref{Fig1}b.

Beyond the non-interacting {\it e-h} pair states, the above symmetry analysis remains valid for the exchange-free exciton states. Figure \ref{Fig2}a shows the calculated energy band dispersions of exchange-free exciton, $E_{S,\VEC{k}_{ex}}^{X(0)}$, of a WSe$_2$-ML with the $\VEC{k}_{ex}$ along $\overline{\Gamma _{ex}M _{ex,2}}$ and $\overline{\Gamma _{ex}K_{ex}'}$ directions, solved from the exchange-free BSE including the direct part of Coulomb interaction only. The lowest exchange-free exciton band over the full BZ is presented in Figure \ref{Fig1}e and \ref{Fig1}f, where $\Gamma_{ex}$, $Q_{ex}/Q_{ex}'$ and $K_{ex}/K_{ex}'$ are identified as the major low-lying excitonic valleys in a WSe$_2$-ML.
In the absence of EHEI, the energy bands of the lowest exciton doublet along $\overline{\Gamma _{ex}M _{ex,2}}$ does remain degenerate while the ones along $\overline{\Gamma _{ex}K_{ex}'}$ are shown valley-split.
Figure~\ref{Fig2}b presents the energy splitting of the lowest exchange-free spin-like exciton doublet, $\Delta _{+ -} ^{(0)} (\VEC{k}_{ex}) \equiv E_{+,\VEC{k}_{ex}}^{X (0)} - E_{-,\VEC{k}_{ex}}^{X (0)}$ (where the subscript $+/-$ indicates the upper/lower band, and the superscript (0) indicates the absence of EHEI), as a function of $\VEC{k}_{ex}$ over the BZ, indeed showing $\Delta _{+ -} ^{(0)} (\VEC{k}_{ex})=0$ (magenta lines) along the three $\overline{\Gamma _{ex}M _{ex,i}}$ paths.

\begin{figure}[!ht]
\includegraphics[width=0.9\columnwidth]{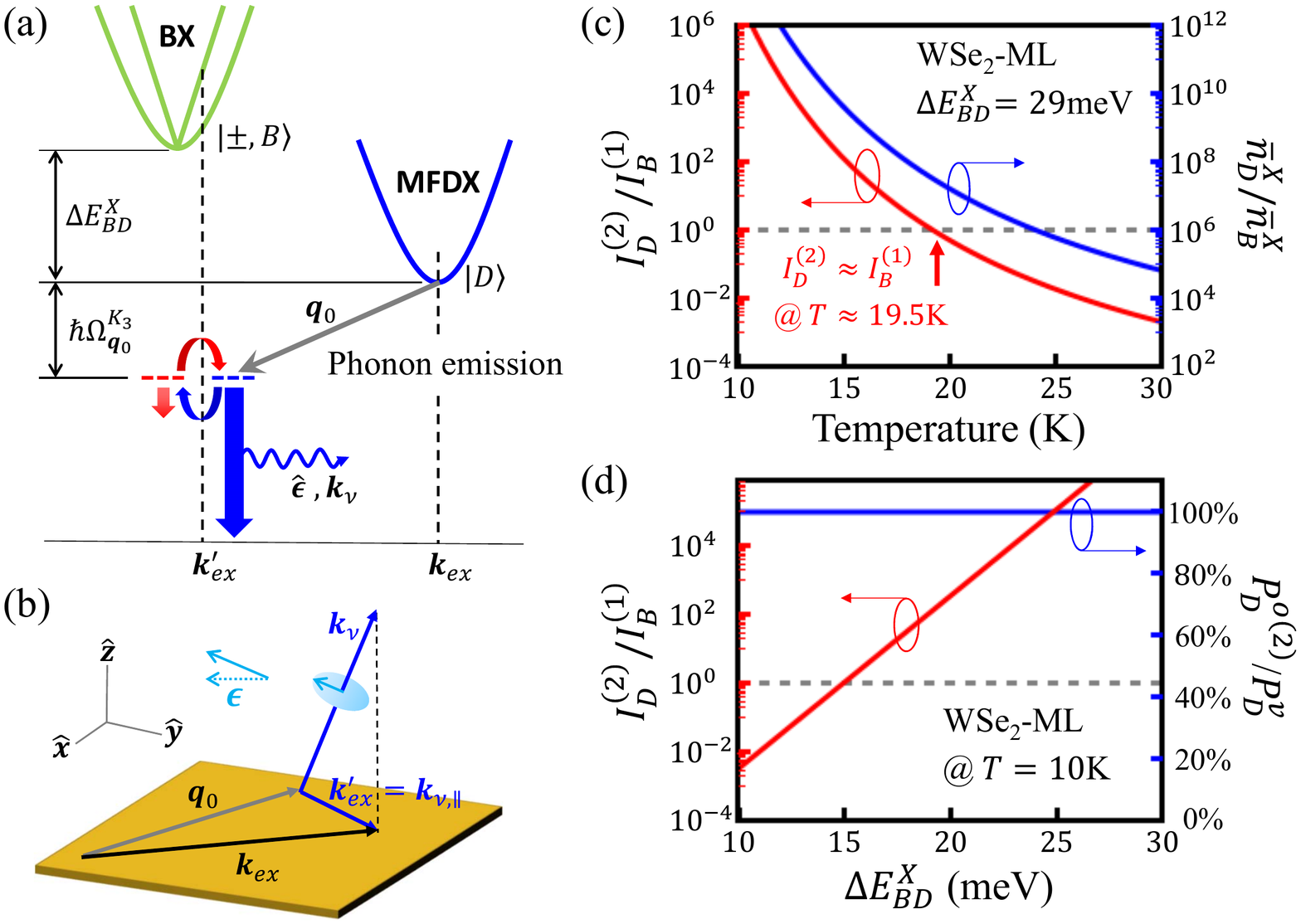}
\caption{(a) Illustrative schematics of a phonon-assisted indirect PL process from a MFDX state $|D\rangle$ and mediated by a BX state $|B\rangle$, and 
(b) the wave vectors of the involved exciton, phonon and photon under the law of momentum conservation. (c) Blue curve: the ratio of the thermal population in $|D\rangle$ and that in $|B\rangle$ as a function of temperature $T$. Red curve: the intensity ratio of the second-order indirect PL from $|D\rangle$ and the direct one from $|B\rangle$. (d) Red line: the intensity ratio of the indirect PL to the direct one with varying $\Delta E _{BD} ^{X} = 10-30$meV. Blue line: the predicted near-unity degree of valley-to-optical polarization conversion for the inter-valley MFDX, $P _{D} ^{o (2)} / P _{D} ^{v}$.}	
 \label{Fig3}
\end{figure}

Figure~\ref{Fig2}d shows the calculated exciton bands of the WSe$_2$-ML with the {\it full} consideration of the both {\it e-h} direct and exchange interactions. Under the effect of EHEI, the exciton bands along the $\overline{\Gamma _{ex}M _{ex,i}}$ paths no longer remain degenerate, as seen in Fig.~\ref{Fig2}e.
For an valley-mixed exciton state written as $|S,\VEC{k}_{ex}\rangle = \sum_{\tau=K,K'}\alpha_{S , \VEC{k}_{ex}}^{\tau}|\tau, \VEC{k}_{ex}\rangle $, in terms of the well-specified valley components, $\{ |\tau,\VEC{k}_{ex}\rangle \}$, the degree of valley polarization (DoV) of the exciton is quantified by $P_{S, \VEC{k}_{ex}}^{v}\equiv \frac{|\alpha_{S , \VEC{k}_{ex}}^{K}|^2-|\alpha_{S , \VEC{k}_{ex}}^{K'}|^2}{|\alpha_{S , \VEC{k}_{ex}}^{K}|^2 + |\alpha_{S , \VEC{k}_{ex}}^{K'}|^2} $.
As seen in Fig.~\ref{Fig2}f, the exciton states lying on the $\overline{\Gamma _{ex}M _{ex,i}}$ paths under the impact of EHEI are featured with $P_{-, \VEC{k}_{ex}}^{v}\sim 0$, while the MFDX ones apart from the $\overline{\Gamma _{ex}M _{ex,i}}$ paths, like those around $Q_{ex,i}/Q_{ex,i}'$ and $K_{ex}/K_{ex}'$ valleys, retain the high DoV of $\mid P_{-, \VEC{k}_{ex}}^{v} \mid \lesssim 100\% $.
Hereafter, we shall pay the main attention on the lowest spin-allowed MFDX states at $K_{ex}/K_{ex}'$ (or named by inter-valley $KK'/K'K$ exciton) whose optical signatures featured with the atractive high degree of polarization were observed in recent cryognic PL measurements. \cite{SFShi2019, WYao2020a}


For the analysis of the phonon-assisted indirect PL, we consider the extended exciton-photon-phonon system with the Hamiltonian $\hat{H} = \hat{H} _X + \hat{H} _{\nu} + \hat{H} _{ph}+ \hat{H} _{X-\nu} + \hat{H}_{X-ph}$, where $\hat{H} _X = \sum _{S \VEC{k} _{ex}} E _{S,\VEC{k}_{ex}}^X \hat{X}_{S,\VEC{k}_{ex}}^{\dagger} \hat{X}_{S,\VEC{k}_{ex}}$ stands for the single-exciton Hamiltonian, $\hat{X}$ ($\hat{X}^{\dagger}$) is the operator annihilating (creating) an exciton, $\hat{H} _{\nu} = \sum _{\VEC{\epsilon}} \sum_{\VEC{k} _{\nu}} \hbar \omega _{ \VEC{k} _{\nu}} ^{\VEC{\epsilon}} \hat{a} _{\VEC{\epsilon}, \VEC{k} _{\nu}} ^\dagger \hat{a} _{\VEC{\epsilon}, \VEC{k} _{\nu}}$ ($\hat{H} _{ph} = \sum_{\lambda} \sum _{\VEC{q}} \hbar \Omega _{\VEC{q}} ^{\lambda} \hat{b} _{\lambda, \VEC{q}} ^\dagger \hat{b} _{\lambda, \VEC{q}}$) is the Hamiltonian of photon (phonon) reservoir, $\omega_{\VEC{k}_{\nu}}^{\VEC{\epsilon}}$ is the frequency of the $\VEC{\epsilon}$-polarized photon with the wave vector $\VEC{k}_{\nu}$, $\Omega_{\VEC{q}}^{\lambda}$ is the frequency of the $\lambda$-kind phonon with the wave vector $\VEC{q}$,\cite{KWKim2014} and $\hat{a} / \hat{a} ^\dagger$ ($\hat{b} / \hat{b} ^\dagger$) are the particle operators that annihilate/create a photon (phonon).
$\hat{H} _{X-\nu}$ ($\hat{H} _{X-ph}$) represents the Hamiltonian of exciton-photon (exciton-phonon) interaction in terms of the $\VEC{\epsilon}$($\lambda$)-dependent coupling constants $\eta_{S,\VEC{k}_{ex}}^{\VEC{\epsilon} , \VEC{k} _{\nu}}$ ($g_{S',\VEC{k}_{ex}';S,\VEC{k}_{ex}}^{\lambda , \VEC{q}}$) (See Ref.\cite{Supplement} for the explicit expressions).

In the process of indirect PL from a WSe$_2$-ML, an exciton initially in the lowest inter-valley MFDX state $|S=-,\VEC{k}_{ex}=\VEC{K}_{ex}/\VEC{K}_{ex}' \rangle \equiv |D\rangle$ is virtually transferred to the intermediate BX states, $|S ^{\prime} , \VEC{k} _{ex} ^{\prime} \in L.C. \rangle \equiv | S ^{\prime}, B \rangle$, assisted by the phonon with $\VEC{k}_{ex} - \VEC{k}_{ex}'\equiv \VEC{q}_0$ and then, from $| S ^{\prime} , B \rangle$, spontaneously emits a photon of wave vector $\VEC{k}_{\nu}$ whose in-plane projection matches the 2D-exciton wave vector, $\VEC{k}_{\nu , \parallel}=\VEC{k}_{ex}'$ (See Fig.\ref{Fig3} for schematic illustration). For simplicity, throughout this work we focus on the specific kind of phonons that makes most the couplings between the inter-valley MFDX states and intra-valley BX ones, which was shown to be the $K_3$ mode in Ref.\cite{WYao2020a,CHLui2020}, i.e. $\lambda = K _{3}$, and consider only the lowest BX doublet, $|S ^{\prime} = \pm, B \rangle$, as the major intermediate states that are intrinsically valley-mixed by the EHEI $\tilde{\Delta}_{KK'}$ and split by $\Delta_{+-}(\VEC{k}_{ex}')=2|\tilde{\Delta}_{KK'}(\VEC{k}_{ex}')|$.

From the Fermi's golden rule (as detailed in Ref.\cite{Supplement}), we show that the averaged transition rates ($\overline{\gamma}_{D}^{(2)}$ and $\overline{\gamma}_{B}^{(1)}$) of the polarization-unresolved indirect PL from $|D\rangle $ and the direct PL from $| \pm , B\rangle$ are explicitly related by
$\overline{\gamma}_{D}^{(2)} / \overline{\gamma}_{B}^{(1)}  = \pi (\frac{a_0}{\lambda})^2 \left| \frac{\hbar g_{BD} \sqrt{N}}{ \Delta E_{BD}^X + \hbar \Omega_{\VEC{q} _0} ^{K _{3}}} \right|^2 $, where $a_0=0.3316$nm is the lattice constant of WSe$_2$-ML \cite{SJCheng2019}, $\lambda \approx 750$nm is the wavelength of the emitted light from $1s$ exciton, $N$ is the total number of primitive cells of the material, $\hbar g_{BD}=\frac{12.3}{\sqrt{N}}$meV is evaluated as the effective exciton-phonon coupling between $|D\rangle$ and $|B\rangle$,\cite{Supplement,KWKim2014} $\Delta E _{BD}^X \equiv E_B^X-E_D^X \approx 29$meV, and $\hbar \Omega_{\VEC{q} _0}^{K _{3}}\approx 26$meV \cite{WYao2020a}.
With the use of those parameters, we count $\overline{\gamma}_{D}^{(2)} / \overline{\gamma}_{B}^{(1)} \gtrsim 10^{-8} $, indicating an extremely slow rate of the indirect PL transition.
The transition rate of indirect PL rate is so low since only a very small portion of intermediate exciton states in the $\VEC{k}_{ex}$ Brillouin zone that are bright to yield the PL, which is measured by the area ratio of the light-cone and the entire zone, i.e.  $\left(\frac{a_0}{\lambda}\right)^2=(\frac{k_c}{| \VEC{G} |})^2$, where $| \VEC{G} | =\frac{2\pi}{a_0}$ and $k_c=\frac{2\pi}{\lambda}$.

Note that the intensity of a PL from an exciton state is determined by the transition rate as well as the exciton population therein, i.e. $I_{B/D}^{(1)/(2)} \propto {\bar \gamma_{B/D}}^{(1)/(2)} \overline{n}_{B/D}^X(T)$, where $\overline{n} _{B/D}^X(T) \propto e^{-E _{B/D} ^{X} /k_B T}$ follows the Boltzmann statistics.
For WSe$_2$-ML, the lowest $K_{ex}/K_{ex}'$ MFDX states are so much lower than the BX ones by $\Delta E _{BD}^X \approx 29$meV and host the tremendously high exciton population at low temperature. Figure \ref{Fig3}c presents the ratio of $\overline{n} ^{X} _{D}/ \overline{n} ^{X}_{B}$ of a WSe$_2$-ML as a function of $T$, showing $\overline{n} ^{X} _D/\overline{n} ^{X} _B \gtrsim 10 ^8$ and $I_{D}^{(2)}/I_{B}^{(1)}>1$ at low temperatures $T<19.5$K. This accounts for the observed pronounced indirect PL peaks from the inter-valley MFDXs in WSe$_2$-MLs in cryogenic PL experiments. \cite{SFShi2019, CHLui2020, WYao2020a}

At last, let us examine the optical polarization, $P_{D}^{o(2)}\equiv \frac{I_{D}^{(2)}(\VEC{\epsilon}_{+})- I_{D}^{(2)}(\VEC{\epsilon}_{-})}{I_{D}^{(2)}(\VEC{\epsilon}_{+})+ I_{D}^{(2)}(\VEC{\epsilon}_{-})}$, of the indirect PL from the inter-valley valley-mixed MFDX state, $|D\rangle = \alpha_{D}^{K}|K,\VEC{k}_{ex}\rangle + \alpha_{D}^{K'}|K',\VEC{k}_{ex}\rangle$. In the second-order perturbation theory (as detailed in Ref.~\cite{Supplement}), we show 
that the optical polarization and the valley polarization of the initial MFDX state of indirect PL are related by 
 \begin{equation}
P _{D}^{o(2)}  \approx P _{D} ^{v} \left( 1 - 4 \Re \left[ \widetilde{\beta}_{BD}^{K K'} \!\! \left( \VEC{k} _{\nu , \parallel} \right) \left( \alpha _{D} ^{K} \right)^{*} \alpha _{D} ^{K'} \right] \right)\, .
\label{Pconvert}
\end{equation}
The last term in Eq.\ref{Pconvert} containing $\widetilde{\beta}_{BD}^{K K'} \equiv - \frac{\widetilde{\Delta}_{K K'} \left( \VEC{k}_{\nu , \parallel} \right)}{ \Delta E_{BD} ^{X} + \hbar \Omega _{\VEC{q}_{0}} ^{K _{3}} }$ arises from the EHEI in the intermediate BX states, through which the $K$($K'$)-valley component in $|D\rangle$ is corss-converted and the optical polarization could be degraded. However, the small value of $|\widetilde{\beta}_{BD} ^{K_{\pm} K_{\mp}} \left( \VEC{k} _{\nu , \parallel} \right)| \sim 10^{-2}$ is estimated with $|\widetilde{\Delta}_{K_{\pm} K_{\mp}}|\sim 1$meV and $( \Delta E_{BD} ^{X} + \hbar \Omega _{\VEC{q}_{0}} ^{K _{3}})\approx 55$meV.  This reveals the native suppression of exchange-induced depolarization in the second-order PL process and accounts for the experimentally observed high degree of polarizations in the phonon-assisted indirect PLs from WSe$_2$-MLs \cite{WYao2020a, SFShi2019}. Figure \ref{Fig3}d presents the calculated $P_D^{o(2)}/P_D^{v}$ for the indirect PL from the MFDX state with varying $\Delta E_{BD}^X$, showing the near-unity degree of the valley-to-optical polarization conversion.


In conclusion, we present a comprehensive theoretical investigation of the full-zone band structures and the momentum-dependent valley polarizations of neutral excitons in TMD-MLs by means of DFT-based numerical computations and symmetry analysis.
Our studies reveal that inter-valley MFDXs, unlike the well recognized BXs, in TMD-MLs are inherently well immune from the exchange-induced valley depolarization under the enforcement of the $D_{3h}$ crystal symmetry. Thus, the valley polarization of the inter-valley $K_{ex}/K_{ex}'$ ($KK'/K'K$) MFDX states are predicted superiorly high and, importantly, fully transferable to the optical polarization of the resulting phonon-assisted indirect PLs. Those findings shed light on the prospective of the valley-based photonics with the utilization of those long-lived, optically accessible and highly valley-polarized inter-valley finite-momentum dark excitons.

P.Y.L. and S.J.C. thank M. Bieniek and P. Hawrylak for fruitful discussion.
This study is supported by the Ministry of Science and Technology, Taiwan, under contracts, MOST 109-2639-E-009-001 and 109-2112-M-009 -018 -MY3, and by National Center for High-Performance Computing (NCHC), Taiwan.

\bibliography{full-zone-exciton-band-refs}{}

\end{document}